\newcommand{\gmn}{g_{\mu\nu}}

\documentclass[twocolumn,nofootinbib,amsmath,amssymb,floatfix]{revtex4}

\usepackage{graphicx}
\usepackage{bm}

\begin{document}

\title{United Dipole Field}

\author{Vladimir F. Tamari}
\email{vftamari@zak.att.ne.jp}
\thanks{This paper was completed in November 1993 but was
not submitted for publication anywhere. The 1993 paper and figures
are presented here with no changes, however necessary (except in
the last sentence in Sec.~\ref{sec:dipolegrav} and the deletion of
the related equation). The paper is now being revised and
expanded. The fact that Einstein's equations have a solution for
gravity waves emitted by a quadrupole but not a dipole does not
alter the conclusions of this paper since a quadrupole is a linear
addition of two dipoles. The help of D.~de Lang in the technical
aspects of this online publication is gratefully acknowledged.}
\affiliation{4-2-8-C26 Komazawa, Setagaya-ku, Tokyo, Japan
154-0012}

\begin{abstract}
The field of an electromagnetic (E) dipole has been examined using
general relativistic (R) and quantum mechanical (Q) points of
view, and an E=Q=R equivalence principle presented whereas the
curvature of the electromagnetic streamlines of the field are
taken to be evidence of the distortion of spacetime, and hence of
the presence of a gravitational field surrounding the dipole.
Using a quasi-refractive index function $N$, with the streamlines
and equipotential surfaces as coordinates, a new dipole
relativistic metric is described, replacing Schwarzschild's for a
point mass. The same principle equates the curvature and other
physical features of the field with fundamental quantum concepts
such as the uncertainty principle, the probability distribution
and the wave packet. The equations of the dipole field therefore
yield the three fields emerging naturally one from the other and
unified without resorting to any new dimensions. It is speculated
whether this model can be extended to dipolar matter-antimatter
pairs.
\end{abstract}


\date{\today}

\maketitle

\section{Introduction}
The quest for a unified theory of Electromagnetism (E)
Gravitational Relativity [R] and Quantum Mechanics (Q) is hampered
by the abstractness of quantum concepts, compared to the other
two. Feynman, in saying that there is no physical `machinery'
causing quantum effects~\cite{feynmanIII} accepts Born's
probabilistic interpretation of quantum events~\cite{born26}. This
lack of physical reality had always bothered Einstein, who
advocated that a successful unification entails ``starting all
over again''~\cite{pais82}. Superstring unification
theory~\cite{schwarz87} does indeed start afresh, but is not yet a
complete theory, and there is still room to search for different
starting points.

The present paper grew out of research in diffraction
theory~\cite{tamari87,tamari03a}, and no complete unification
theory was attempted or is of course claimed from this elementary
treatment. Nevertheless, it was found that certain physical
features of the electromagnetic dipole field could have intriguing
quantum probabilistic interpretations. And when it was discovered,
through general relativistic analysis of the streamline curvature,
that the dipole might also possess a gravitational field, the
question arose whether the elements of a prototype unified field
theory were now at hand.

The diffraction of a dipole-photon is examined in
Section~\ref{sec:dipolediff} as a more exact form of the
Huygens-Fresnel principle. In Section~\ref{sec:dipolegrav} the
equipotential surfaces $\phi$, and the orthogonal streamlines $S$
of a dipole field are proposed as the coordinates of a $\gmn$
relativistic metric symmetric around the dipole axis and plane
only, replacing the sphero-symmetric Schwarzschild metric for a
particle: the severe curvature of the streamlines in the dipole
origin is interpreted as an indication of an unexpectedly powerful
gravitational field there. This field is characterized by a radial
quasi-refractive index function $N(R,\theta)$ of the distortion of
spacetime. In Section~\ref{sec:dipoleqfield} the dipole electric
field is shown to coincide with the Gaussian probability function
while the dipole's frequency and amplitude attenuations follow
those of a quantum wave packet. An E=Q=R equivalence principle is
stated in Section~\ref{sec:equivalence}, and it is speculated in
Section~\ref{sec:conclusion} whether the dipole unified field
model may be adapted to other particles than the photon, perhaps
through the concept of virtual pairs of matter-antimatter, and
whether the strong force could be attributed to a dipole
gravitational field.

\section{Electromagnetic Dipole Diffraction}\label{sec:dipolediff}
Using intuitive hydrodynamical arguments, Tamari has proposed that
an expanding photon field must have both a foreword linear and a
radial component~\cite{tamari90}, resembling the bow wave of a
boat. Miller~\cite{miller91} has shown that Maxwell's equations
can yield a more exact form of the Huygens-Fresnel principle for a
point source, replacing it by a dipole with a given
spatio-temporal relation of the phase and separation $d$ of
positive and negative charges $a$, emitting a field with
wavelength $\lambda_0=2\pi/k$, and with a potential field:

\begin{equation}\label{eq:dipolefld}
\phi=\frac{ad}{4\pi R}\left[ k\sin (kR)(1+\cos\theta) +\cos (kr)
\frac{\cos\theta}{R}\right]
\end{equation}

Both the bow-wave and the spatio-temporal dipole models of the
photon have a characteristic geometry of expanding nested
near-circular equipotential surfaces joined at the origin. For the
sake of simplifying the analysis here, an approximate dipole field
is taken from the literature~\cite{feynmanII} whereby, for a
dipole of moment $D=ad/4\pi$ aligned with the $z$-axis and
symmetrical about the origin, with a time harmonic phase term $H$,
the potential is:

\begin{equation}\label{eq:appdipolefld}
\phi=\frac{DH\cos\theta}{R^2}
\end{equation}

\noindent with orthogonal streamlines:

\begin{equation}\label{eq:streamlines}
S=\frac{R}{\sin^2 \theta}
\end{equation}

\noindent as shown in Figure~1. These approximations also omit the
magnetic and quadrupole components. While
Eq.~\ref{eq:appdipolefld} and Eq.~\ref{eq:streamlines} are
symmetrical about the $x-y$ plane, both the bow-wave and the
spatio-temporal dipole field of Eq.~\ref{eq:dipolefld} vanish in
the $-z$ regions.

The curvature of the dipole streamlines $S$, along which the
electromagnetic energy flows, is also a feature seen in Braunbek
and Laukien's~\cite{braunbek52} analysis of Sommerfeld's rigorous
solution of diffraction around an infinite
half-plane~\cite{sommerfeld96}. In Tamari's Streamline Diffraction
Theory~\cite{tamari87} the various curved diffracting streamlines
inclined at angles $-\frac{\pi}{2}\leq\theta \leq +\frac{\pi}{2}$
carry all the Fourier components of the field~\cite{goodman68} in
a fan-shaped flow pattern. Along the $+z$ axis the dipole
streamline is straight and carries the electromagnetic field at a
speed $c$, with equiphasals separated by the wavelength
$\lambda_0$. But what of all the other curved $S$?

\begin{figure}[htb]
\includegraphics[width=8.6cm]{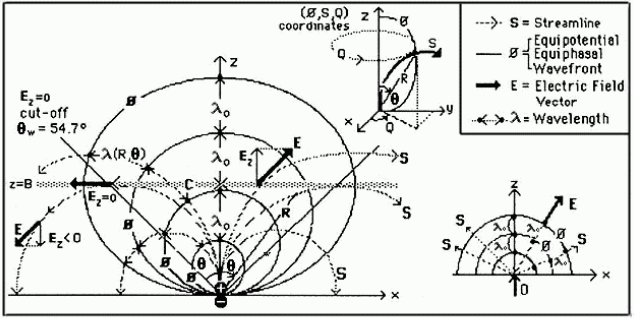}
\caption{\label{fig:1}Diagram of a dipole field. On the right, a
Huygens-Fresnel wavelet. }
\end{figure}

\begin{figure}[htb]
\includegraphics[width=8.6cm]{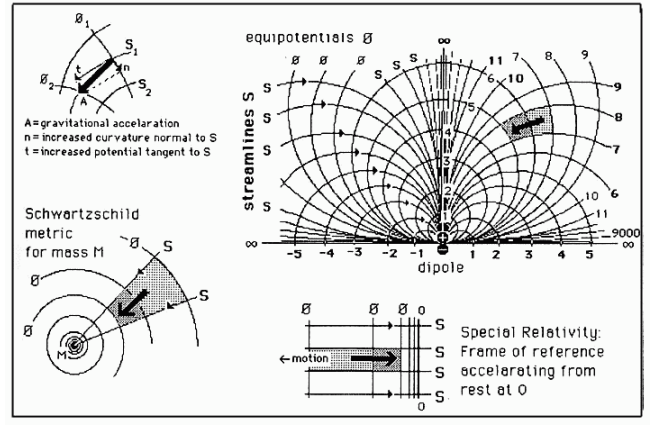}
\caption{\label{fig:2} Gravitational acceleration $\rightarrow$
acts along streamlines ($S$) towards increasing potential
($\phi$): its tangential and normal components point towards
increasingly curved space-time.}
\end{figure}

\section{Relativistic Dipole Gravitation}\label{sec:dipolegrav}
By definition no energy flows between streamlines, and each curved
$S$ can be considered a unique pathway carrying light. But
according to Einstein's general theory of relativity it is the
presence of a gravitational field that causes a ray of light to
bend~\cite{einstein88}. Postponing for the present the question of
what the source of the dipole's gravitational field would be, the
geometry of the streamlines themselves is considered. Since they
carry light, $S$ must be \textit{geodesics} of the field,
\textit{i.e.} following a minimum path through the dipole's
spacetime~\cite{schutz85}. Figure~2 illustrates the intuitive
basis for regarding the dipole field as a gravitational field. A
new $(\phi,S,Q,t)$ coordinate system based on
Eq.~\ref{eq:appdipolefld} and Eq.~\ref{eq:streamlines} and
symmetric about the $z$-axis is therefore defined, whereby, for a
time $t=0$, the distance element is:
\begin{equation}\label{eq:distelement}
ds^2=d\phi^2+ dS^2 + R^2 \sin^2 \theta dQ^2 - \frac{c^2 dt^2}{N}
=\gmn x_1^2 x_2^2
\end{equation}

\begin{eqnarray}\label{eq:gmn}
ds^2=& & \nonumber\\
&\left[\frac{1}{\sin^4 \theta}+\frac{4 D^2 H^2 \cos^2
\theta}{R^6}\right]dR^2 &\dots\dots g_{11}\nonumber\\
+&\left[\frac{4D^2 H^2 \cos\theta\sin\theta}{R^6}-\frac{4 R
\cos\theta}{\sin^5\theta}\right]dRd\theta &\dots g_{12},~g_{21}\nonumber\\
+&\left[\frac{4R^4 \cos^2\theta}{\sin^6\theta}+\frac{D^2
H^2\sin^2 \theta}{R^2}\right]d\theta^2 &\dots\dots g_{22}\nonumber\\
+&\left[R^2\sin^2 \theta\right]dQ^2 &\dots\dots g_{33}\nonumber\\
+&\left[\frac{c^2R^3}{DH\sqrt{3\cos^2 \theta +1}}\right]dt^2
&\dots\dots g_{44}\nonumber\\
& &
\end{eqnarray}

\noindent with all the other $\gmn=0$. The $N$ term is to be
explained at length below. A full treatment using tensor algebra
proving that the $\gmn$ obey Einstein's equations for general
relativity, and finding the curvature tensor
$g^{\mu\nu}G_{\mu\nu}$ although essential, is beyond the scope of
this paper. Fortunately, there is a mathematically simple and
exact device, which helps conceive the curvature of spacetime in
vivid physical terms.

In dynamics the curvature of the path of a particle is evidence of
acceleration, and hence a change in local velocity.  According to
Einstein, ``A curvature of rays of light can only take place when
the velocity of propagation of light varies with
position''~\cite{einstein61}, a rather problematic statement
because of the constancy of c in his relativity theory.  But as
Eddington~\cite{eddington90} explained, it is the ``coordinate
velocity'' $v$ of light, which diminishes in the presence of a
gravitational field, and hence a quasi-refractive index of
spacetime $N$ can be defined where:

\begin{equation}\label{eq:N}
  N=\frac{c}{v}
\end{equation}

Since the local velocity defines $N$, it is a measure of
\textit{both} space contraction and time dilation, the more
conventional relativistic terms. Eddington, Lodge, and
others~\cite{israel92} used this concept of a variable density of
spacetime, a gradient-index field $N$, akin to that in geometrical
optics~\cite{bornwolf} to describe relativistic phenomena such as
the properties of a black hole.

The use of $N$ transforms a relativistic coordinate system into a
classical one: At first $S$ are considered ``straight'' geodesics
in the vacuum surrounding the dipole. Alternately, the same $S$
can be considered the curved rays of light in a glass-like medium
with a gradient index of refraction function $N(R,\theta)$. While
the two treatments in no way change the physical realities of the
field, a variable speed of light $v$ can now be unambiguously
defined, and a geodesic can therefore be considered the solution
to the eikonal equation of geometrical optics~\cite{bornwolf}:

\begin{equation}\label{eq:N2}
N^2=(\nabla \phi)^2
\end{equation}

In the absence of masses or charges, $N=1$, but in gravitational
fields light `slows down' and $N>1$, reaching infinity at the edge
of a black hole.  In dynamics, the curvature $k$ of the path of a
moving particle is related to its acceleration
$A=(dv/dt)\bm{t}+kv^2 \bm{n}$, where $\bm{t}$ and $\bm{n}$ are
unit vectors tangent and normal to the path
respectively~\cite{wylie83}. In the present case, $A$ becomes the
\textit{gravitational acceleration}, with its well-known
relativistic dependence on the curvature of space, as illustrated
in Figure~2. The streamline curvature $k$ is derived from the
standard equation for line curvature, and from
Eq.~\ref{eq:streamlines}:

\begin{equation}\label{eq:curvature}
k=\frac{R^2+2\left(\frac{dR}{d\theta}\right)^2-R\left(\frac{d^2R}{d\theta^2}\right)}
{\left(\left(\frac{dR}{d\theta}\right)^2+R^2\right)^{\frac{3}{2}}}=
\frac{3\left(1+\frac{2}{\tan^2\theta}\right)}{R\left(1+\frac{4}{\tan^2\theta}\right)^{\frac{3}{2}}}
\end{equation}

Or, from a geometrical optical relation for a ray in a gradient
index field, $|k|=\bm{n}\nabla \log N$~\cite{bornwolf3}. In
Figure~3, rays curve towards regions of increasing $N$, which can
be found directly from Eq.~\ref{eq:N2} and
Eq.~\ref{eq:appdipolefld}:

\begin{equation}\label{eq:Nd}
N_d= \frac{DH\sqrt{3\cos^2 \theta +1}}{R^3}
\end{equation}

The path of any geodesic can now be traced through radial
$d\theta$ segments of the $N$ field with ease and precision by
using Snell's law of refraction

\begin{equation}\label{eq:Snell}
N_1 \sin \theta_1 =N_2 \sin \theta_2
\end{equation}

\noindent as shown in Fig~3. In making these calculations
Eq.~\ref{eq:Nd} cannot be applied as it is because successive
$N_1$, and $N_2$ segments share the same $R$ but different angles.
Therefore, from Eq.~\ref{eq:appdipolefld} and Eq.~\ref{eq:Nd}, and
putting $H=\phi=D=1$ since they cancel out in Eq.~\ref{eq:Snell}:

\begin{equation}\label{eq:Ndnorm}
N_d=\frac{\sqrt{3\cos^2\theta+1}}{(\cos \theta)^{\frac{3}{2}}}
\end{equation}
\noindent (normalized value for N calculations)

\begin{figure}[htb]
\includegraphics[width=8.6cm]{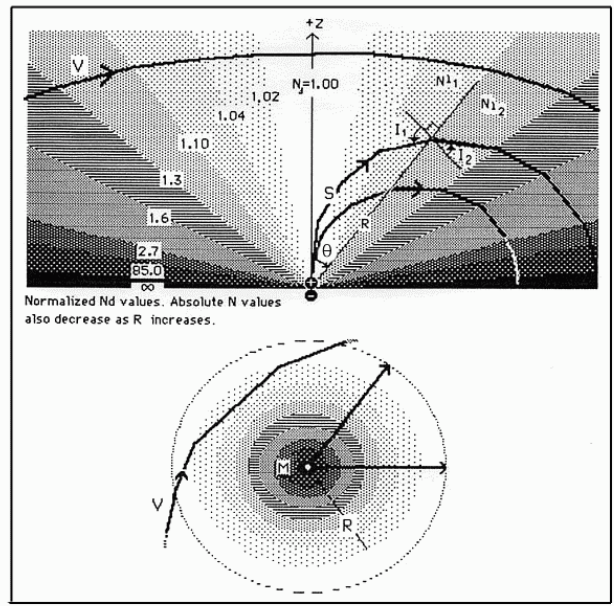}
\caption{\label{fig:3} Schematic representation of the path
streamlines $S$ and independent rays $V$ through the
quasi-refractive index fields $N_d(R,\theta)$ of a dipole (top)
and in the Schwartschild $N_s(R)$ metric around a mass $M$
(bottom). Typical refractive angles $\theta_1$ and $\theta_2$ show
bending of light through increments of the $N$ field.}
\end{figure}

\begin{figure}[htb]
\includegraphics[width=8.6cm]{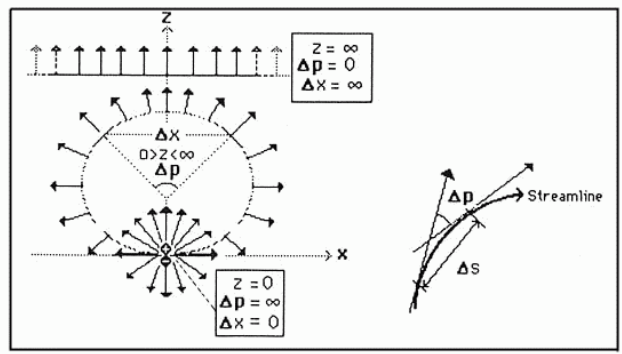}
\caption{\label{fig:4} Physical basis of the uncertainty of
position $\Delta x$ and momentum $\Delta p$ in the dipole field
and along a streamline.}
\end{figure}

Since a dipole's gravity causes the surrounding space to curve,
bending its own streamlines, then any \textit{other} particle or
ray will similarly experience the same gravitational field. Based
on this intuitive but unproven assumption, Eq.~\ref{eq:Snell} and
Eq.~\ref{eq:Ndnorm} were used to trace the path of various rays
$V$ initiated at given points $P(R,\theta)$ and angles of
incidence $I$ within the $N$ field. The path of the streamlines
$S$ of Eq.~\ref{eq:streamlines}, passing through a point $P$ was
confirmed if fine angular increments $<0.1\deg$ were used to
calculate the refracted rays.

But what of the Schwarzschild~\cite{schutz2} metric:
\begin{equation}\label{eq:Schwarzmetric}
ds^2=\frac{dR^2}{F}+R^2d\theta^2-Fc^2dt^2
\end{equation}

\noindent with $F=(1-2GM/c^2R)$, where $G$ is the constant of
universal gravitation? Here, instead of a dipole at the origin,
there is a mass $M$ and space-time is distorted only radially.
Putting $ds=0$, the condition for a geodesic, and $d\theta=0$
since the field is spherically symmetric, an expression for the
local `velocity' $v=dR/dt$ is obtained. And using Eq.~\ref{eq:N},
this give a quasi-refractive index function $N_sR$ for this
metric~\cite{eddington90}:

\begin{equation}\label{eq:Ns}
N_s=\frac{1}{F}
\end{equation}

\noindent indicating the `density' of space-time due to a single
\textit{point mass} at the origin, the expression for a point
charge being similar. Substituting Eq.~\ref{eq:Ns} back into
Eq.~\ref{eq:Schwarzmetric}, this result can be generalized for all
metrics: Space contracts by a factor of $\sqrt{N}$ along
streamlines (due to the change in potential) and time dilates by
$\sqrt{N^{-1}}$. But in the dipole case $N$ was \textit{derived}
from the $(\theta,S)$ coordinates, and does not need to be
included back in, which justifies Eq.~\ref{eq:distelement}.

The asymmetry of the dipole metric, compared to Schwarzschild's is
due to the presence of \textit{two opposite charges} at the
origin. Another difference is that $N_s =\infty$ at the origin,
but $N_d$ has no singularities anywhere. Very near the charges,
the dipole potential, and hence the $N_d$ value gets quite
complicated, and no longer follows that of
Eq.~\ref{eq:appdipolefld}. In general, the local $N$ value at a
point $P$ is obtained by the the vector addition of the all
electric fields at $P$  caused by surrounding dipoles in different
positions and orientations.

\section{The Dipole Quantum Field}\label{sec:dipoleqfield}
Some of the dipole field's physical features will now be
interpreted from a quantum mechanical point of view. But as
Hawking has remarked, ``[in quantum mechanics] the unpredictable,
random element comes in only when we try to interpret the wave in
terms of the positions and velocities of particles. But maybe that
is our mistake: maybe there are no particle positions and
velocities, but only waves.''~\cite{hawking88} And Schr\"{o}dinger
always had reservations on the probabilistic interpretation of
quantum mechanics.~\cite{scott67} Adopting these views is
necessary here to justify the following physical interpretations:

\subsection{The particle-wave duality}
The dipole moment $D$ is a quantum quantity since it depends on
the distance between the charges and their quantified strength,
and therein lies the `particle' aspect of the field under
consideration. The `wave' aspect of the field is simply the
time-harmonic classical dipole field itself. This in no way
disputes the results of quantum mechanics, only the `causes'. For
example, an intensity distribution can be said to be the
probabilistic accumulation of many whole particle photons. Here
the photon will be described as a single continuous classical wave
with local intensity fluctuations, which then cause random
particle events during absorption or emission, and occurring
\textit{within the sensor}.

\subsection{The uncertainty principle}
Heisenberg himself cited diffraction~\cite{heisenberg49} as one
illustration of his uncertainty relations. But it will be argued
here that \textit{uncertainty relations exist precisely because
waves diffract}. As in Figure~4, at the origin, the dipole field
is basically concentrated in one point, so $\Delta x=0$, but the
streamline directions carrying the field's momentum $p$ point in
an infinite number of directions, hence $\Delta p=\infty$. Very
far from the origin, the streamlines are basically parallel to the
$+z$ axis and $\Delta p=0$, but the wavefront has spread very
widely and $\Delta x=\infty$. At intermediate points $\Delta x$
and $\Delta p$ can be mathematically related in several ways. For
example, along any single streamline, $\Delta p$ amounts to the
curvature the streamline (when the streamline is straight, $\Delta
p=0$), while $\Delta x$ is the distance element along the
streamline. The physical basis of Planck's constant $h$ in the
relation $\Delta x \Delta p \geq h$ is elusive, but must be sought
in the relation between charge quanta and the geometry of the
dipole field at the origin.

\subsection{The probability function}
It is well known that $-\nabla \phi= E$, the electric field
strength, the vectorial version of $N$. Using an alternative
derivation~\cite{feynmanI} for the dipole's electric field
intensity parallel to the $z-$axis, again with $D=H=1$:

\begin{equation}\label{eq:Ez}
E_z=\frac{3\cos^2 \theta -1}{R^3}
\end{equation}

Putting $3\cos^2\theta=1$ gives a cut-off angle $T_w= 54.73\deg$
beyond which $E_z = 0$ and there is no foreword momentum, as shown
in Figure~1, where:

\begin{equation}\label{eq:c}
C=B \tan T_w=1.4139 B
\end{equation}

In Figure~5, the $E_z$ values (normalized by an amplitude factor
of $2.363 B^{-2}$) of the dipole field along $z=B$, was compared
to the Gaussian probability distribution, related to the quantum
probability function~\cite{brandt85}.

\begin{equation}\label{eq:prob}
P(x)=\frac{1}{j\sqrt{2\pi}}e^{-\frac{x^2}{2j^2}}
\end{equation}

Where $j=C/3=0.4713 B$,  (the standard deviation, a third of the
significant width before the curve dwindles asymptotically to
zero), was chosen so that the Gaussian fits over $E_z(x)$.
Figure~5 shows that there are only minor differences between
$P(x)$ and $E_z(x)$ for $|x|<|C|$.  For $|x|>|C|$, $E_z$ becomes
negative, and more significantly, no energy reaches a screen at
$B$, since all the streamlines curve down before reaching it. What
are the quantum implications of such a field, particularly the
cut-off angle $C$? The probability amplitude can be directly
related to the curvature of $S$: Increasing the curvature
increases the angle of incidence of $S$ on $B$, and hence the
smaller the $E_z$ vector.

\begin{figure}[htb]
\includegraphics[width=8.6cm]{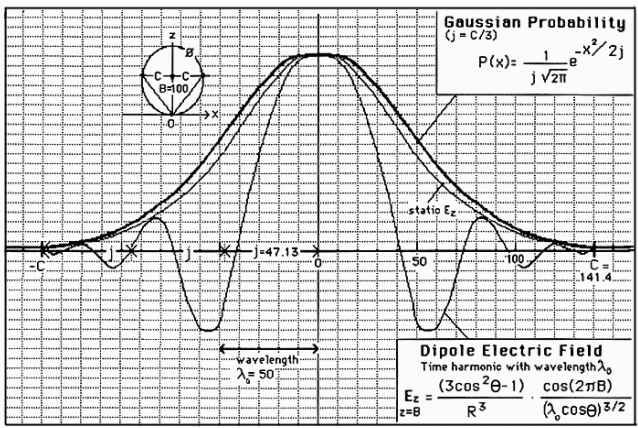}
\caption{\label{fig:5} Physical basis of probability in dipole
field.}
\end{figure}

\subsection{The Wave Packet}
In the case of a time-harmonic dipole, using
Eq.~\ref{eq:appdipolefld} and Eq.~\ref{eq:Nd}, and the fact that
the wavelength is $\lambda_0$ along the $+z$ axis, it was found
that along the line $z=B$:

\begin{equation}\label{eq:wavepacket}
E_{z(t=0)}= \cos \left(\frac{2\pi B}{\lambda_0
(\cos\theta)^{\frac{3}{2}}}\right) \frac{3\cos^2 \theta -1}{R^3}
\end{equation}

\noindent so that the equipotentials occur at ever-decreasing
distances. In quantum mechanics, the photon field is found to
comprise a wave packet, with a spectral function $f(k)$ specifying
the infinitely diminishing wavelengths and amplitudes fitting
within a Gaussian envelope~\cite{brandt85}. It is seen in Figure~5
how $E_{z(t=0)}$ resembles such a function in all details, as to
amplitude and wavelength diminuation: as the wave packet expands,
the equiphasals intersect $B$ and $S$ at ever shorter intervals,
as in Fig.~1: there is a blue-shift in the field until the
wavelength $I(R,\theta)$, and hence $v$, becomes zero on the $x$
axis.

\section{E=R=Q Principle}\label{sec:equivalence}
According to the analysis above, an
electromagnetic-relativistic-quantum-mechanical equivalence
principle can now be generalized as follows: ``in the field
surrounding a system of electromagnetic charges, the gravitational
field is equivalent to the curvature of the diffracting
streamlines, while the local quantum mechanical state is
equivalent to the gradient of the potential.''

This of course has only been examined in the single case cited
above, the first-order dipole approximation of
Eq.~\ref{eq:appdipolefld}. But other configurations produce curved
streamlines, such as two point emitters of like charge,
diffraction through an aperture in an opaque screen, diffraction
from any continuous distribution of sources, and others.

\section{Conclusion}\label{sec:conclusion}
The electromagnetic dipole field has been examined from
relativistic and quantum mechanical points of view, with
preliminary evidence that it makes up a unified field in which an
E=Q=R equivalence principle can be asserted: The three fields here
appear to emerge, one from the other, in a satisfying and elegant
way, and without the need for any new dimensions. Can Dirac's
oppositely charged virtual pairs of particles
(matter-antimatter)~\cite{gamov66} be considered a kind of dipole,
with electromagnetic, gravitational and quantum fields such as
those discussed above? Is the atomic strong force the result of
some kind of short-range gravitational field such as the one
modelled above? These are some of the questions that emerge from
the present study.

\begin{acknowledgments}
Dr. D.~A.~B.~Miller's clarification of his dipole formulation, and
stimulating discussions with Miss Mona~Tamari related to the
unified dipole field are gratefully acknowledged.
\end{acknowledgments}



\begin{thebibliography}{99}

\bibitem{feynmanIII}
R.~Feynman, R.~Leighton, and M.~Sands, \textit{The Feynman
Lectures on Physics III}, Addison-Wesley,  1-10 (1965).

\bibitem{born26}
M.~Born, Zeit. Phyz. \textbf{38}, 803 (1926).

\bibitem{pais82}
A.~Einstein, quoted in A.~Pais, \textit{Subtle is the Lord....},
Oxford Univ. Press, 461 (1982).

\bibitem{schwarz87}
J.~Schwarz, ``Superstring Unification'', in \textit{300 Years of
Gravitation}, (S.~Hawking and W.~Israel, Eds.) Cambridge Univ.
Press, 652 (1987).

\bibitem{tamari87}
V.~Tamari, \textit{The Cancellation of Diffraction in Wave
Fields}, Optoelectronics-Devices and Tech. Mita Press \textbf{2},
59-82 (1987).

\bibitem{tamari03a}
V.F.~Tamari, E-print: physics/0303073 (2003).

\bibitem{tamari90}
V.~Tamari, ``Bow-Wave Geometry'', post-deadline paper in
\textit{Huygens' Principle 1690-1990 Conference}, Scheveningen,
The Netherlands (1990).

\bibitem{miller91}
D.~A.~B.~Miller \textit{Huygens' wave propagation principle
corrected}, Optics Lett.~\textbf{16}, 1370-72 (1991).

\bibitem{feynmanII}
R.~Feynman, R.~Leighton, and M.~Sands, \textit{The Feynman
Lectures on Physics II} op.~cit.~II-6-3.

\bibitem{braunbek52}
W.~Braunbek and 0.~Laukien, Optik \textbf{9}, 174-179 (1952); see
also M.~Born and E.~Wolf, \textit{Principles of Optics}, 6th. Ed
Pergamon, 575-77 ().

\bibitem{sommerfeld96}
A.~Sommerfeld, Math.~Ann.~\textbf{47}, 317 (1896); see also
M.~Born and E.~Wolf, \textit{Principles of Optics},
op.~cit.~556-575.

\bibitem{goodman68}
J.~Goodman, \textit{Introduction to Fourier Optics}, McGraw-Hill
(1968).

\bibitem{einstein88}
A.~Einstein, \textit{The Meaning of Relativity}, Princeton Univ.
Press, 93 (1988).

\bibitem{schutz85}
B.~Schutz, \textit{A First Course in General Relativity},
Cambridge Univ. Press, 166 (1985).

\bibitem{einstein61}
A.~Einstein, \textit{Relativity}, Bonanza Books, 76 (1961).

\bibitem{eddington90}
A.~Eddington, \textit{Space, Time and Gravitation}, Cambridge
Univ. Press, 107-109 (1990).

\bibitem{israel92}
W.~Israel, ``Dark stars: The Evolution of an Idea'', in
\textit{300 Years of Gravitation}, op.~cit.~204 (1992).

\bibitem{bornwolf}
M.~Born and E.~Wolf, \textit{Principles of Optics}, op.~cit.~13.

\bibitem{bornwolf2}
M.~Born and E.~Wolf, \textit{Principles of Optics}, op.~cit.~112.

\bibitem{wylie83}
C.~Wylie and L.~Barret, \textit{Advanced Engineering Mathematics},
McGraw Hill, 775 (1983).

\bibitem{bornwolf3}
M.~Born and E.~Wolf, \textit{Principles of Optics}, op.~cit.~124.

\bibitem{schutz2}
B.~Schutz, \textit{A First Course in General Relativity},
op.~cit.~258.

\bibitem{hawking88}
S.~Hawking, \textit{A Brief History of Time}, Bantam, 183 (1988).

\bibitem{scott67}
W.~Scott, \textit{Erwin Schr\"{o}dinger}, Univ.~Mass.~Press
(1967); see A.~Pais, \textit{Subtle is the Lord...},
op.~cit~(Ref.~3 above), p.443f.

\bibitem{heisenberg49}
W.~Heisenberg, \textit{The Physical Principles of the Quantum
Theory}, Dover, 23 (1949).

\bibitem{feynmanI}
R.~Feynman, R.~Leighton, and M.~Sands, \textit{The Feynman
Lectures on Physics}, op.~cit.~I-6-9.

\bibitem{brandt85}
S.~Brandt and H.~D.~Dahmen, \textit{The Picture Book of Quantum
Mechanics}, John Wiley \& Sons, 36 (1985).

\bibitem{gamov66}
G.~Gamov, \textit{Thirty Years That Shook Physics}, Dover, 131
(1966).

\end{thebibliography}
\end{document}